\documentclass[universe,review,accept,pdftex]{Definitions/mdpi}
\usepackage{amsmath}
\usepackage{amsfonts}
\usepackage{amssymb,bm}
\usepackage{siunitx}
\usepackage{color}
\usepackage{braket}
\usepackage{graphics}
\firstpage{1}
\pubvolume{10}
\issuenum{3}
\articlenumber{119}
\pubyear{2024}
\copyrightyear{2024}
\externaleditor{Academic Editor: Gerald B. Cleaver} 
\datereceived{29 January 2024}
\dateaccepted{28 February 2024}
\datepublished{4 March 2024}
\hreflink{https://doi.org/10.3390/\linebreak universe10030119} 
\Title{The Nature of Dark Energy and Constraints on Its Hypothetical Constituents
from Force Measurements }

\TitleCitation{The Nature of Dark Energy and Constraints on Its Hypothetical Constituents
from Force Measurements }

\Author{{Galina L. Klimchitskaya} ${}^{1,2}$*\orcidA{}
and
Vladimir M. Mostepanenko ${}^{1,2,3}$\orcidB{}}

\AuthorNames{Galina L. Klimchitskaya,
Vladimir M. Mostepanenko}

\AuthorCitation{Klimchitskaya, G.L.;  Mostepanenko, V.M.}

\address{%
$^{1}$ \quad Central Astronomical Observatory at Pulkovo of the Russian Academy of Sciences, 196140 Saint Petersburg, Russia\\
$^{2}$ \quad Peter the Great Saint Petersburg
Polytechnic University, 195251 Saint Petersburg, Russia\\
$^{3}$ \quad  {Kazan Federal University}, 420008 Kazan, Russia}

\corres{Correspondence: g.klimchitskaya@gmail.com}

\abstract{This review  considers the theoretical approaches to the understanding
of dark energy which comprises approximately 68\% of the energy of our
Universe and explains an acceleration in its expansion. Following a
discussion of the main approach based on Einstein's equations with
the cosmological term, the explanations of dark energy using the
concept of some kind of scalar field are elucidated. These include the
concept of a quintessence and modifications of the general theory of
relativity by means of the scalar-tensor gravity exploiting the chameleon,
symmetron, and environment-dependent dilaton fields and corresponding
particles. After mentioning several laboratory experiments allowing to
constrain the hypothetical scalar fields modeling the dark energy,
special attention is devoted to the possibility of constraining the
parameters of chameleon, symmetron, and environment-dependent dilaton
fields from measuring the Casimir force. It is concluded that the
parameters of each of these fields can be significantly strengthened in
near future by using the next generation setups in preparation suitable
for measuring the Casimir force at larger separations.}


\begin{document}
\section{Introduction}

The concept of expanding Universe, which goes back to the Friedmann
solutions of Einstein's equations published in 1922 \cite{1} and
1924 \cite{2}, assumes that its expansion should decelerate with
time due to the gravitational attraction of both visible and dark
matter. It was a big surprise when, analyzing the redshift data of
supernovae in binary systems, the two research teams independently
found in 1998 that the Universe expansion is accelerating (see the
pioneer Refs. \cite{3,4} and Refs. \cite{5,6} for a review).

If one wishes to explain the acceleration of the Universe expansion
in the framework of the general theory of relativity, it is
necessary to admit that there is an additional form of invisible matter
with a positive energy density $\varepsilon > 0$, as it holds for
both the usual and dark matter, but with a negative pressure,
$P < 0$. Such a matter is characterized by the equation of state
\begin{linenomath}
\begin{equation}
P = -w\varepsilon,
\label{eq1}
\end{equation}
\end{linenomath}
where an acceleration in the expansion holds for $w > 1/3$. This kind
of invisible matter violating the strong energy condition was called
the dark energy.

The advent of dark energy would not be so unusual if it constituted a
small fraction of the total energy of the Universe. It turned out,
however, that if one would like to preserve the standard cosmological
scenario based on the general theory of relativity, the observational
data demand that the dark energy constitutes about 68\% of the Universe
energy \cite{5,6}. When it is considered that the dark matter
contributes approximately 27\% of the Universe energy, only 5\% remain
for the visible, baryonic, matter.

There are many theoretical approaches to the understanding of the
nature of dark energy. These approaches can be grouped into the four
main divisions. The first of them describes the dark energy by means
of the cosmological constant $\Lambda$ introduced into equations of
the general theory of relativity by Einstein \cite{7} for other
purposes.

The second group of approaches to the description of dark energy
considers it as some kind of classical time-varying scalar field
called a quintessence. The cosmological applications of similar
fields were considered in Refs. \cite{8,9}, whereas the term
a quintessence was introduced in Ref. \cite{10}.

The third group of approaches allows any change in the action and
equations of the general theory of relativity by combining the
metrical tensor with the classical scalar field within the formalism
of scalar-tensor gravity in order to make the concept of dark energy
unnecessary \cite{11}. The chameleon field, symmetron field, and the
environment-dependent dilaton field were used in the literature for
this purpose. Some of these approaches dispense with the need for
either the dark energy or dark matter (see, e.g., Ref. \cite{12}).
The modifications of the gravitational theory are also allowed in
the unified models of dark matter and dark energy introducing the
so-called dark fluid \cite{13}.

Note that the main ideas of the above three groups of approaches
can be considered as based on the concepts of classical physics
although quantum physics was used in their further developments.
As to the approaches of the fourth group aiming to understand the
nature of dark energy, they consider it as composed of some
hypothetical elementary particles with unusual physical preperties
which give rise to the negative pressure. The most popular particles
of such kind are the chameleons, which possess a variable mass
depending on the density of matter in the environment \cite{14,14a}.
Another candidate for a dark energy particle is the symmetron whose
interaction constant with the usual matter depends on the
environmental density \cite{15,16,17}. There are also other
hypothetical particle candidates for the role of constituents of
dark energy, e.g., the environment-dependent dilaton \cite{18}.
The classical fields with the variable masses and interaction
constants were introduced in the third group of approaches mentioned
above, whereas the unusual particles, such as chameleons, symmetrons
etc., are the result of their quantization.

In this review, we compare the approaches from the above four groups
by the level of their credibility and discuss the main particle candidates
for the role of dark energy constituents. Next, we pass to the constraints
on the parameters of chameleon, symmetron, and environment-dependent dilaton
fields following from different laboratory experiments. The main attention
is paid to the constraints which can be obtained from measuring the Casimir
force arising between the closely spaced macroscopic bodies due to the
zero-point and thermal fluctuations of the electromagnetic field.

There are also many alternative attempts to solve the dark energy issue,
which are listed below for completeness. Thus, one can mention suggestions
to consider modified gravity theories that introduce additional degrees of
freedom in the gravitational and/or matter action \cite{45,19a}.
It was also suggested to phenomenologically modify the Friedmann equation
by additional terms that depend on the matter density in a nonlinear way
\cite{19b,19c,19d}. Another option considered in the literature is to alter
the mass-energy evolution equation with bulk viscosity terms \cite{19e,19f,19g}.

Alternatively, some authors believe that dark energy may be only an apparent effect.
They hypothesize that the supernovae data may be biased if the observer is located
in a local underdense region (see, e.g., Ref.~\cite{19h}) or suppose that the
supernovae sources tend to be associated with overdensities (see, e.g.,
Ref.~\cite{19i}). Finally, many papers focus on the role of matter inhomogeneities
and anisotropies that  may affect the cosmic expansion due to backreaction or
statistical sampling effects (see, e.g., Refs.~\cite{19j,19k,19l,19m,19n}).

The review is organized as follows. In Section 2, the theoretical
approaches to understanding of the physical nature of dark energy
based on classical physics are briefly considered and compared.
Section 3 is devoted to a discussion of different particle
candidates for the role of constituents of dark energy. The already
obtained laboratory constraints on the parameters of chameleon,
symmetron, and environment-dependent dilaton fields, as well as the
prospective constraints obtainable from force measurements,
including the Casimir force, are presented in Section 4. Section 5
contains the discussion, and in Section 6 the reader will find our
conclusions.

Below the relativistic units are used with $c = \hbar = 1$, where $c$
is the speed of light and $\hbar$ is the reduced Planck constant.

\section{Approaches to Theoretical Description of Dark Energy Based on
Classical Physics}

As discussed in Section~1, there are three groups of such kind approaches to
understanding of what the dark energy is and none of them is either excluded
or finally confirmed.

We begin with probably the most common approach describing the accelerations
in the Universe expansion on the basis of classical Einstein equations with the
cosmological term
\begin{linenomath}
\begin{equation}
R_{ik}-\frac{1}{2}Rg_{ik}-\Lambda g_{ik}=8\pi GT_{ik},
\label{eq2}
\end{equation}
\end{linenomath}
where $R_{ik}$ is the Ricci tensor, $R$ is the scalar curvature of the space-time,
$\Lambda$ is the cosmological constant, $g_{ik}$ is the metrical tensor,
$G$ is the gravitational constant, and $T_{ik}$ is the stress-energy tensor of
both the visible and dark matter.

Equation (\ref{eq2}) provides a very plausible explanation for the dark energy
because in the homogenous isotropic 3-space of expanding Universe the metrical
tensor is diagonal. Thus, raising the index $k$ in Eq.~(\ref{eq2}) and
rearranging the cosmological term to the right-hand side of this equation,
one obtains
\begin{linenomath}
\begin{equation}
R_{i}^{\,k}-\frac{1}{2}R\delta_{i}^{\,k}=8\pi G\left(T_{i}^{\,k}+
\frac{\Lambda}{8\pi G}\delta_{i}^{\,k}\right),
\label{eq3}
\end{equation}
\end{linenomath}
where  $\delta_{i}^{\,k}$ is the Kronecker symbol. From this equation it is
seen that the effective stress-energy tensor caused by the cosmological constant is
\begin{linenomath}
\begin{equation}
{T_{(\Lambda)i}}^{k}=
\frac{\Lambda}{8\pi G}\delta_{i}^{\,k}.
\label{eq4}
\end{equation}
\end{linenomath}

Taking into account that in the homogeneous isotropic space for the stress-energy
tensor of any kind of matter it holds \cite{19}
\begin{linenomath}
\begin{equation}
T_0^{\,0}=\varepsilon, \qquad T_{1}^{\,1}=T_{2}^{\,2}=T_{3}^{\,3}=-P,
\label{eq5}
\end{equation}
\end{linenomath}
where $ \varepsilon$ is the energy density and $P$ is the pressure, one obtains
from Eq.~(\ref{eq4}) the energy density, pressure and equation of state of the
dark energy resulting from the cosmological constant
\begin{linenomath}
\begin{equation}
\varepsilon_{\Lambda}=\frac{\Lambda}{8\pi G},\qquad
P_{\Lambda}=-\frac{\Lambda}{8\pi G},\qquad
P_{\Lambda}=-\varepsilon_{\Lambda}.
\label{eq6}
\end{equation}
\end{linenomath}

Thus, in this case, Eq.~(\ref{eq1}) is satisfied with $w=w_{\Lambda}=1$ in
violation of the second inequality of the strong energy condition
\begin{linenomath}
\begin{equation}
\varepsilon+P\geqslant 0,  \qquad \varepsilon+3P\geqslant 0
\label{eq6a}
\end{equation}
\end{linenomath}
valid for the usual
and dark matter.

In spite of the fact that Eq.~(\ref{eq2}) belongs to the classical physics,
it has long been understood \cite{20} that the leading divergent term in the vacuum
expectation values of the stress-energy tensor of quantized fields has the same
geometric form as the cosmological term
\begin{linenomath}
\begin{equation}
\langle0|T_{ik}|0\rangle=I_{\infty}g_{ik},
\label{eq7}
\end{equation}
\end{linenomath}
where $I_{\infty}$ is an infinitely large constant. This is valid in both
the Minkowski space-time and in the curved background of expanding Universe
\cite{21,22} as can be seen, for instance, by the method of dimensional
regularization \cite{23}.

From Eqs.~(\ref{eq5}) and (\ref{eq7}), it follows that
\begin{linenomath}
\begin{eqnarray}
&&
\langle 0|T_0^{\,0}|0\rangle=\varepsilon_{\rm vac}=I_{\infty},
\nonumber\\
&&
\langle 0|T_1^{\,1}|0\rangle=\langle 0|T_2^{\,2}|0\rangle=
\langle 0|T_3^{\,3}|0\rangle=-P_{\rm vac}=I_{\infty},
\label{eq8}
\end{eqnarray}
\end{linenomath}
i.e., the equation of state of the quantum vacuum
\begin{linenomath}
\begin{equation}
P_{\rm vac}=-\varepsilon_{\rm vac}
\label{eq9}
\end{equation}
\end{linenomath}
is the same as due to the cosmological constant in Eq.~(\ref{eq6}).

Thus, the vacuum stress-energy tensor of quantized fields could offer a plausible
explanation for a generation of the cosmological constant. However, the great
difficulty, called the vacuum catastrophe \cite{24}, arises from the infinitely
large values of $I_{\infty}$, $\varepsilon_{\rm vac}$ and $P_{\rm vac}$.
Even if one makes a cutoff in the expression for $I_{\infty}$ at the Planck
momentum, the obtained energy density is of the order
\begin{linenomath}
\begin{equation}
\varepsilon_{\rm vac}\sim 10^{111}~\mbox{J/m}^3.
\label{eq9a}
\end{equation}
\end{linenomath}
At the same time, the observed acceleration in the Universe expansion demands
the value of the cosmological constant in Eq.~(\ref{eq2})
\begin{linenomath}
\begin{equation}
\Lambda\approx 10^{-52}~\mbox{m}^{-2}.
\label{eq9b}
\end{equation}
\end{linenomath}
This results in the corresponding
value of the vacuum energy density
\begin{linenomath}
\begin{equation}
\varepsilon_{\Lambda}=\frac{\Lambda}{8\pi G}\sim 10^{-9}~\mbox{J/m}^3,
\label{eq9c}
\end{equation}
\end{linenomath}
which is different by the factor of $10^{120}$ from the estimation of
$\varepsilon_{\rm vac}$ in Eq.~(\ref{eq9a})
obtained from quantum field theory \cite{6,25}.
In Ref.~\cite{23}, it was suggested to consider the value of
$\Lambda$ from Eq.~(\ref{eq9b}) as a renormalized value of the
cosmological constant as opposed to the enormously large bare value
\begin{linenomath}
\begin{equation}
\Lambda_{\rm vac}=8\pi G\varepsilon_{\rm vac}\sim 10^{68}~\mbox{m}^{-2}.
\label{eq9d}
\end{equation}
\end{linenomath}
Some grounds for such an approach are given by the quantum field theory in curved
space-time \cite{21,22}, but the rigorous justification could be reached only
in the framework of quantum theory of gravitation which is not yet available.

In spite of this problem, the cosmological constant, whose value is determined
experimentally like the values of all other fundamental constants, provides
a pretty convincing explanation for the acceleration in the Universe expansion.
In fact Eq.~(\ref{eq2}) including the cosmological term can be considered as
entirely classical with no connection with the problem of quantum vacuum.
As a result, the cosmological constant is commonly considered as one of the
main elements of the standard cosmological model Lambda-CDM along with the
cold dark matter formed by the nonrelativistic particles (axions, weakly
interacting massive particles) and the usual barionic matter.

The second group of approaches to an explanation of the acceleration in the
Universe expansion considers the dark energy as a time-varying classical scalar
field $\Phi$ called the quintessence \cite{8,9,10}. Unlike the dark energy
described by the cosmological constant, where the quantity $w$ in Eq.~(\ref{eq1})
is constant, $w=1$, for the quintessence $w$ depends on the form of the field
potential $V(\Phi)$ and may vary with time.

There are many models of the quintessence dark energy proposed in the literature
(see, for instance, Refs.~\cite{26,27,28,29,30,31,32,33} and review \cite{34})
using different forms of the potential $V(\Phi)$ \cite{10,27,28,34,35,36,37,38,39,40}.
Typically the sum of the actions of the general theory of relativity and
the quintessence
field is chosen in the form
\begin{linenomath}
\begin{equation}
S=\int\! d^4x\sqrt{-g}\left[\frac{1}{16\pi G}\,R-\frac{1}{2}g^{ik}\,
\frac{\partial\Phi}{\partial x^i}\,\frac{\partial\Phi}{\partial x^k}
-V(\Phi)\right],
\label{eq10}
\end{equation}
\end{linenomath}
where $g$ is the determinant of the metrical tensor and the interaction with the
usual baryonic matter $\psi$ is lacking. Because of this, the total action is the
sum of $S$ and the action of the baryonic matter $S_m[\psi]$.

In the space-time of expanding Universe the quantity $w$ takes the form \cite{34}
\begin{linenomath}
\begin{equation}
w\equiv w_{\Phi}=
\frac{2V(\Phi)-\left(\frac{\partial\Phi}{\partial t}\right)^2}{2V(\Phi)+
\left(\frac{\partial\Phi}{\partial t}\right)^2}.
\label{eq11}
\end{equation}
\end{linenomath}

It was shown that with the exponential potential \cite{26,27,34}
\begin{linenomath}
\begin{equation}
V(\Phi)=V_q(\Phi)=V_0e^{-\lambda\sqrt{8\pi G}\Phi},
\label{eq12}
\end{equation}
\end{linenomath}
where $\lambda={\rm const}$, the equation of state of the quintessence dark energy
approaches to Eq.~(\ref{eq1}) with $w=w_q=1-\lambda^2/3$.
As a result, the quintessence approach to the dark energy becomes capable to make
approximately the same theoretical predictions for the accelerated expansion of the
Universe as the standard model using the cosmological constant.

Note also that in some models of a quintessence the quantity $w$ defined in Eq.~(\ref{eq1})
satisfies the inequality $w>1$. This means that the kinetic energy of a quintessence field
is negative leading to a catastrophic acceleration of the Universe expansion without bounds.
As a result, the distances between individual particles, even inside an atom, go to infinity.
In the literature, this is called the Big Rip caused by the phantom energy \cite{40a}.
There are also models of kinetic quintessence with a nonstandard form of negative kinetic
energy but $0<w<1$ \cite{40b}. The fact is worth mentioning that the concept of a
quintessence field is used for a solution of the so-called coincidence problem, i.e., why
the energy densities of dark matter and dark energy are of the same order of magnitude in
the present epoch of cosmic history \cite{28} (see also Refs.~\cite{20,25}).

The third group of theoretical approaches essentially based on the classical physics
admits modifications of the general theory of relativity in such a way that an
introduction of the dark energy could be obviated. The most well known modification
of the general theory of relativity is the scalar-tensor theory which assumes that
the gravitational interaction is determined by the combined action of the metrical
tensor and the scalar field $\Phi$ (see the pioneer paper \cite{41}, reviews
\cite{42,43} and the monograph \cite{44}).

The typical action of the scalar-tensor theory is the sum of the action defined in
Eq.~(\ref{eq10}) and the action of usual matter, $S_m$, which is, however,
coupled with the field $\Phi$ in this case
\begin{linenomath}
\begin{equation}
S_{\rm int}=S_{\rm int}[A^2(\Phi)g_{ik},\psi],
\label{eq13}
\end{equation}
\end{linenomath}
where $A(\Phi)$ is some function describing the coupling to matter. Thus, in the
Brans-Dicke theory \cite{41}
\begin{linenomath}
\begin{equation}
A(\Phi)=A_{\rm BD}(\Phi)=e^{-\frac{\sqrt{\pi G}}{C}\Phi},
\label{eq14}
\end{equation}
\end{linenomath}
where $C={\rm const}$.

Due to Eq.~(\ref{eq13}) the effective potential depends on the usual matter.
For example, for the dust-like matter with an energy density $T_0^{\,0}=\varepsilon$
and $P=0$ one has \cite{11}
\begin{linenomath}
\begin{equation}
V_{\rm eff}(\Phi)=V(\Phi)+\varepsilon, \qquad
\Box\Phi=\frac{\partial V_{\rm eff}(\Phi)}{\partial\Phi}.
\label{eq15}
\end{equation}
\end{linenomath}

Both the potential $V(\Phi)$ and the function $A(\Phi)$ take different forms in
various models proposed in the literature \cite{45}. Thus, the chameleon field with
a choice \cite{14,14a,46}
\begin{linenomath}
\begin{eqnarray}
&&
V(\Phi)=V_{\rm ch}(\Phi)=\frac{M^{4+n}}{\Phi^n}, \qquad
A(\Phi)=A_{\rm ch}(\Phi)\approx 1+C\sqrt{8\pi G}\Phi,
\nonumber\\
&&
V_{\rm eff}(\Phi)=V_{\rm eff,\,ch}(\Phi)=V_{\rm ch}(\Phi)+
C\sqrt{8\pi G}\varepsilon\Phi,
\label{eq16}
\end{eqnarray}
\end{linenomath}
where $M$ is a parameter with the dimension of mass, $n$ is an integer number,
$C$ is a constant of the order of unity, is used in the models of dark energy.
The effective mass of chameleon field is larger in the regions of larger
density, $m_{\Phi}^2\sim \varepsilon^{(n+2)/(n+1)}$.

Another choice used in the models of dark energy is the symmetron field for
which \cite{16,17}
\begin{linenomath}
\begin{eqnarray}
&&
V(\Phi)=V_{\rm s}(\Phi)=-\frac{m^2}{2}\Phi^2+\frac{\lambda}{4}\Phi^4, \qquad
A(\Phi)=A_{\rm s}(\Phi)\approx 1+\frac{\Phi^2}{2M^2},
\nonumber\\
&&
V_{\rm eff}(\Phi)=V_{\rm eff,\,s}(\Phi)=V_{\rm s}(\Phi)+
\varepsilon A(\Phi),
\label{eq17}
\end{eqnarray}
\end{linenomath}
where  $\lambda$ is the dimensionless constant of self-interaction and $m$ is one
more parameter with the dimension of mass. The coupling strength of the symmetron
field to the usual matter is of the order of $\Phi/M$. It is perceptible in the
regions of low density $\varepsilon/M^2\ll m^2$ and goes to zero in the regions
of sufficiently high density $\varepsilon/M^2>m^2$ \cite{47}.

Another class of modifications of the general theory of relativity
replaces the standard action of this theory linear in $R$ with a nonlinear
one \cite{44,48}
\begin{linenomath}
\begin{equation}
S=\frac{1}{16\pi G}\int\!d^4x\sqrt{-g} \,f(R) +\int\!d^4x\sqrt{-g}{\cal L}_M,
\label{eq18}
\end{equation}
\end{linenomath}
where ${\cal L}_M$ is the Lagrangian density of the usual matter,
$f(R)$ can be presented as a series expansion
\begin{linenomath}
\begin{equation}
f(R)=\ldots +\frac{\beta_{-2}}{R^2}+\frac{\beta_{-1}}{R}+f(0)+R+\beta_2R^2+\ldots\,,
\label{eq18a}
\end{equation}
\end{linenomath}
and $f(0)=2\Lambda$ is expressed via the cosmological constant.

As shown in Ref.~\cite{48}, the function of the form $f\sim 1/R^n$ with $n>0$
in Eq.~(\ref{eq18}) can explain the observed acceleration in the Universe expansion.
It was shown, however, that the theories described by the action (\ref{eq18}) are
in fact the versions of the scalar-tensor theories of gravity considered above
\cite{49,50}. Thus, the dynamically equivalent to (\ref{eq18}) action written in
terms of an additional scalar field $\chi$ is
\begin{linenomath}
\begin{equation}
S=\frac{1}{16\pi G}\int\!d^4x\sqrt{-g}\left[f(\chi)+f^{\prime}(\chi)(R-\chi)\right]
 +\int\!d^4x\sqrt{-g}{\cal L}_M.
\label{eq18b}
\end{equation}
\end{linenomath}

Really, the variation of this action with respect to $\chi$ results in the equation
of motion
\begin{linenomath}
\begin{equation}
f^{\prime\prime}(\chi)(R-\chi)=0,
\label{eq18c}
\end{equation}
\end{linenomath}
where $f^{\prime}(\chi)=\partial f(\chi)/\partial\chi$. This means that $\chi=R$ if
$f^{\prime\prime}(\chi)\neq 0$ and Eq.~(\ref{eq18b}) reduces to Eq.~(\ref{eq18}).

Next, by introducing one more scalar field $\Phi=f^{\prime}(\chi)$, one can transform
the action (\ref{eq18}) to the action of a Brans-Dicke theory with the potential
\cite{50}
\begin{linenomath}
\begin{equation}
V(\Phi)=\chi(\Phi)\Phi-f\left(\chi(\Phi)\right).
\label{eq18d}
\end{equation}
\end{linenomath}

This means that any constraints obtained for a chameleon or symmetron fields from
measuring the Casimir force (see Section 4) can be reformulated as the corresponding
constraints on the function $f^{\prime}(R)$ known as the scalaron field or,
alternatively, as the cosmological scalar field in theories of modified $f(R)$
gravity. The latter, however, is outside the scope of this review.

A comprehensive review of these and many others theories of modified
gravity and their applications to cosmology is given in Ref.~\cite{45}.

As is seen from the above, both the second and third groups of approaches to the
theoretical description of an acceleration in the Universe expansion are heavily
based on the consideration of some hypothetical scalar field whose form of potential,
the function describing an interaction with matter
and some parameters are not fixed uniquely. In this sense, the first
approach exploiting the cosmological term in Einstein's equations seems preferable
because it operates with only one parameter, the cosmological constant, which can be
considered as a fundamental constant like the electric charge, speed of light,
Planck constant etc. In the next section we discuss what could be added to this
situation by the quantum theory, which brings an interpretation of the classical
scalar fields used in the models considered above in terms of particles.

\section{Particle Candidates for the Role of Constituents of Dark Energy}

As discussed in previous section, the classical chameleon and symmetron fields were
introduced in the context of modified gravity. This makes their immediate quantization
problematic because the consistent quantum theory related to the standard part of
gravitation described by the metrical tensor is not yet available. For this reason,
the action of the form of Eq.~(\ref{eq10}) or the sum of Eqs.~(\ref{eq10}) and
(\ref{eq13}) cannot be directly presented in the operator form.

It is possible, however, to consider the action of a scalar field $\Phi$ and its
interaction with the matter fields separately of the gravitational action containing
the scalar curvature. In so doing, the metrical tensor in the action (\ref{eq13}),
describing an interaction of the matter fields with $\Phi$, is understood as the usual
function in the spirit of quantum field theory in curved space-time \cite{21,22}.

Using this approach, the chameleon field can be quantized and the resulting particles
are called the chameleons. Then it is possible to consider the interaction of
chameleons with the curved gravitational background and with the elementary particles
of the Standard Model. Thus, the quantum corrections to the chameleon potential were
investigated in Ref.~\cite{51}. The effect of production of chameleons from vacuum in
the early Universe was considered in the linear approximation in Ref.~\cite{52}
by the method of Bogoliubov transformations. It was shown that in the radiation
dominated Universe this effect makes a strong impact on the Universe evolution.

In addition to interaction with the baryon particles, chameleons can be coupled to
photons via the additional term of the form $\Phi F_{ik}F^{ik}$, where $F_{ik}$ is
the tensor of the electromagnetic field. This term is in fact the linear approximation
to the exact interaction which contains the chameleon field in the exponent \cite{53}
\begin{linenomath}
\begin{equation}
S_{\rm int,\,ch}=-\frac{1}{4}\int\!d^4x\,e^{\frac{\Phi}{M}}F_{ik}F^{ik},
\label{eq19}
\end{equation}
\end{linenomath}
where $M$ is a fictitious mass controlling the coupling strength of chameleons to photons.
Due to the interaction (\ref{eq19}), chameleons can be turned to photons and vice versa
in an external magnetic field.

Similar situation also holds as to the quantization of the symmetron field. If one considers
its action separately from the action of gravitation, the symmetron field can be
quantized with the metrical tensor $g_{ik}$ being a classical function. The resulting quanta
are called symmetrons. As discussed in Section~2, the coupling of symmetron field to the
usual baryonic matter vanishes if the local energy density is large enough and is restored
in the regions with sufficiently low energy density.

On the classical level, the symmetron field does not interact with the electromagnetic
field. However, in the framework of quantum field theory, it was shown that quantum
corrections generate the interaction Lagrangian density between symmetrons and photons
of the form \cite{47,54}
\begin{linenomath}
\begin{equation}
{\cal L}_{\rm s}=\frac{\Phi^2}{M^2}\,A_{\rm s}^{-4}g^{ik}g^{ln}F_{il}F_{kn},
\label{eq20}
\end{equation}
\end{linenomath}
where $M$ is some new energy scale and $A_{\rm s}=A_{\rm s}(\Phi)$ is defined in
Eq.~(\ref{eq17}). This is the so-called axion-like coupling.

One more particle with unusual physical properties, which can be considered as a
hypothetical constituent of dark energy, is the environment-dependent dilaton.
The dilaton scalar field and its associated particles arise in many theoretical
approaches beyond the Standard Model, e.g., in the extra-dimensional theories with
a varied volume of compactified space, in the scalar-tensor theories of gravity,
in string theory etc. (see, e.g., Refs.~\cite{44,55,56,57}).

Below we consider the model of an environment-dependent dilaton field which
is formulated in the context of scalar-tensor gravity. In fact this field combines
the properties of the quintessence, chameleon, and symmetron fields. Thus, similar to
the chameleon and symmetron fields, it is described by the sum of actions defined in
Eqs.~(\ref{eq10}) and (\ref{eq13}). The function $A$ describing the coupling of
an environment-dependent dilaton to matter is of the same form as was discussed
for symmetrons in Eq.~(\ref{eq17}) \cite{58,59}
\begin{linenomath}
\begin{equation}
A_{\rm d}(\Phi)=1+\frac{A_0}{2M^2}(\Phi-\Phi_0)^2,
\label{eq21}
\end{equation}
\end{linenomath}
where $\Phi_0$ is the current value of the dilaton field and $A_0$ is a constant.

As to the dilaton potential, it takes the exponential form \cite{58,59} like for
the quintessence field [see Eq.~(\ref{eq12})]
\begin{linenomath}
\begin{equation}
V_{\rm d}(\Phi)=V_0e^{-\lambda\sqrt{8\pi G}\Phi},
\label{eq22}
\end{equation}
\end{linenomath}
as opposed to the power-type potentials (\ref{eq16}) and (\ref{eq17}) for the
chameleon and symmetron fields, respectively.

In the regions of space with sufficiently high density of matter, it holds
$\Phi\approx\Phi_0$ and the coupling of the dilaton field to matter becomes
negligibly small, although in the regions with low density the coupling of
the dilaton field to matter is of the order of gravitational strength.
In this regard the environment-dependent dilaton behaves in the same way as the
symmetron. Similar to chameleons, however, the effective mass of a dilaton
increases with increasing density of the environment.

The quantization of the  environment-dependent dilaton field can be performed
under the same conditions as discussed above for the chameleon and symmetron fields.
In addition to coupling with baryons, the dilaton particles can be coupled to
photons. This coupling has the form of Eq.~(\ref{eq19}), the same as for chameleons
\cite{60}.

\section{Constraints on the Particle Constituents of Dark Energy
from Force Measurements}

The hypothetical scalar fields (the chameleon, symmetron, and
environment-dependent dilaton) discussed in Sections~2 and 3 interact with the usual
matter and can be constrained in the laboratory experiments in a number of ways.
Thus, it was shown \cite{61} that individual atoms inserted into large high-vacuum
chamber do not screen the chameleon field and the force acting on them from this
field can be measured by means of atom interferometry.

One more method for searching chameleon particles uses their interaction with the
electromagnetic field. For observation of oscillations between the chameleon and
photon states, the vacuum chamber was used where the magnetic field of 5~T was
initiated \cite{53}. As a result, in the plane (effective
chameleon mass)$\,\times\,$(coupling to photon parameter),
rather large region was excluded.

Strong limits on the parameters of chameleons were placed also by means of the
gravity resonance spectroscopy used to measure the quantum states of ultracold
neutrons confined near a mirror \cite{62}. These limits are by the five orders
of magnitude stronger than the previously known ones obtained from spectroscopic
measurements \cite{63}.

The same methods can be used for searching and constraining the symmetrons and
environment-dependent dilatons. For instance, in Refs.~\cite{64,65} it was shown
that the parameters of symmetrons can be constrained by means of atom interferometry.
As one more example, the possibility to constrain dilatons by measuring the
dilaton-photon conversion in strong magnetic field was considered in Ref.~\cite{66}
(see also the review \cite{67} where several other possibilities are considered).

Constraints on the chameleon, symmetron and dilaton fields and respective particles
can be obtained not only from the laboratory experiments mentioned above but from
astrophysics and cosmology as well. One can mention constraints found from galaxy
clusters thermodynamic profiles, gravitational lensing, and caustic techniques
\cite{67a,67b,67c,67d}. Specifically, the amplitude of the chameleon field and its
coupling strength to matter were constrained by combining the gas and lensing
measurements of the cluster \cite{67a}. The upper limits on the strength of chameleon
force were placed by comparing X-ray and weak lensing profiles of the galaxy clusters
\cite{67b}. It should be noted, however, that the constraints found from astrophysics
and cosmology do not admit an immediate comparison with the laboratory constraints
because the former unavoidably depend on some indefinite factors whereas the latter
are obtained in the fully controlled environments.

Below we concentrate our attention on constraining the parameters of chameleons,
symmetrons, and environment-dependent dilatons which can be obtained from force
measurements at short separations below a few micrometers. The point is that
at such small distances between the material bodies the dominant force is not
the gravitational one, but the Casimir force caused by the zero-point and thermal
fluctuations of the electromagnetic field \cite{68}. Precision measurements of
the Casimir force have long been used for constraining the Yukawa-type
corrections to Newton's law of gravitation and the interaction constant and mass
of axions as the possible constituents of dark matter (see, e.g.,
Refs.~\cite{69,70,71,72} and reviews \cite{72a,73,73a}).

The standard approach to obtaining constraints on some hypothetical force
$F_{\rm hyp}$  from measuring the Casimir force is the following.
According to the experimental data obtained over some separation interval, the
theoretical expression for the Casimir force is confirmed within the total
error  $\Delta F$ which includes the random and systematic experimental errors
as well as possible theoretical uncertainties. The hypothetical force, e.g.,
from the Yukawa-type interaction or due to the axion exchange, is calculated
in the experimental configuration as a function of separation and the parameters
of this interaction. Since the hypothetical force was not observed, its magnitude
is restricted by the inequality
\begin{linenomath}
\begin{equation}
|F_{\rm hyp}(a)|<\Delta F(a),
\label{eq23}
\end{equation}
\end{linenomath}
where $a$ is the value of separation. Then, by analyzing this inequality at
different separations within the measurement interval, the strongest constraints
on the parameters of hypothetical force are obtained \cite{69,70,71,72,73}.

This methodology can also be applied to the possible constituents of dark energy,
such as  chameleons, symmetrons, and environment-dependent dilatons.
The obtained results are considered in the following subsections.

\subsection{Constraints on Chameleons from Measuring the Casimir Force}

The possibility to constrain the chameleon parameters from measuring the Casimir force
was proposed in Refs.~\cite{74,75} and further elaborated in Ref.~\cite{76}.
Thus, in Ref.~\cite{76} the hypothetical force due to the presence of chameleons was
calculated in the configurations of two parallel plates and a sphere above a plate.
The latter configuration was used in all precise experiments on measuring the
Casimir force \cite{68}.

As was noted in Section~2, different forms of the potential $V(\Phi)$ in Eq.~(\ref{eq10})
have been proposed in the literature. The results of Ref.~\cite{76} are obtained with the
most widely used potential of the form of Eq.~(\ref{eq16}) and with the exponential
potential
\begin{linenomath}
\begin{equation}
V_{\rm ch}(\Phi)=\widetilde{\Lambda}_0^4\,e^{\frac{\widetilde{\Lambda}^n}{\Phi^n}}.
\label{eq24}
\end{equation}
\end{linenomath}

The first term in the power expansion of Eq.~(\ref{eq24}) corresponds to the vacuum
energy density required for explanation of the accelerated expansion of the Universe
and the second with $\widetilde{\Lambda}=\widetilde{\Lambda}_0=M$ results in the
potential (\ref{eq16}).

Taking into account that the mass of the chameleon field strongly depends on the
density of the environment, the macroscopic bodies are characterazed by the so-called
thin shells regarding this field \cite{14,14a}. Let the body have
the density $\rho_b$
and outside the body the density of matter is $\rho_m$. Then, deep inside the body,
\begin{linenomath}
\begin{equation}
\Phi\approx\Phi_b\equiv\Phi_{\min}(\rho_b),
\label{eq24a}
\end{equation}
\end{linenomath}
where the effective potential $V_{\rm eff}(\Phi)$ reaches its minimum value at
$\Phi_{\min}$. As to the region outside the body, there it holds
\begin{linenomath}
\begin{equation}
\Phi\approx\Phi_m\equiv\Phi_{\min}(\rho_m).
\label{eq24b}
\end{equation}
\end{linenomath}

According to Ref.~\cite{76}, for the thin-shelled bodies almost all of the change
from $\Phi_m$ to $\Phi_b$ happens in the thin shell near a surface of the body.
It turned out that the hypothetical force due to the presence of chameleon field
between the thin-shelled bodies is much weaker than for sufficiently thin bodies
where the thin shell near the surface is not formed \cite{74,75,76}.

According to the analysis performed in Ref.~\cite{76}, the test bodies used in
measurements of the Casimir force (see Refs.~\cite{68,77} for a review) have the
thin shells for the most realistic models of the chameleon field used in the
literature. It was noted also \cite{76} that if the thin shells in the test bodies
are absent, all the constraints on  Yukawa interaction obtained from measuring
the Casimir force remain valid for the chameleon theories.

In Ref.~\cite{76}, the chameleon force was calculated between two parallel plates and
between a sphere and a plate with account of the effect of thin shells for the
potentials of the form (\ref{eq16}) and (\ref{eq24}). As a result,
rather wide regions were excluded
in the plane (chameleon-to-matter coupling)$\,\times\,$(energy scale of chameleon potential) using the data of the most precise measurements of the Casimir force.
For strengthening of the obtained constraints, it was suggested to perform measurements
of the thermal Casimir force at larger separations and to use larger test bodies in
order to avoid the effect of thin shells which decreases the magnitude of the chameleon
force.

\subsection{Constraints on a Symmetron Field from Measuring the Casimir Force}

As discussed in Section~3, the coupling of the symmetron field to
the barionic matter increases
in the regions of low density and goes to zero with increasing density of matter. This field
and the corresponding particles are described by the sum of actions (\ref{eq10}) and
(\ref{eq13}), where the potential and the function $A(\Phi)$ describing the coupling to
matter are given by Eq.~(\ref{eq17}).

Constraints on the symmetron field following from measurements of the Casimir force can be
obtained using the same methodology as described above in the case of a chameleon field.
One should calculate the additional force caused by the symmetron field in the configuration
of two plates or a sphere above a plate used in the Casimir experiments.  If the theoretical
expression for the Casimir force is confirmed by the measurement data in the limits of
some error, then the magnitude of any additional force is restricted by this error.

In Ref.~\cite{78} the exact analytical solutions for the profiles of a symmetron field were
found in the space near an infinite mirror occupying a semispace and between two such
mirrors separated by a gap. The first of this solutions was applied for calculation of an
additional frequency shift in the experiment measuring the reflection of ultracold neutrons
by a neutron mirror in the gravitational field of the Earth \cite{79,80}.
The second analytical solution concerning the case of two mirrors can be applied for
calculation of the additional force induced by the symmetron field in the proposed
CANNEX experiment on measuring the Casimir force between two parallel plates at
separations up to $10~\upmu$m and more \cite{81,82,83}.
The principal scheme of the setup of this experiment, which is also discussed in Section~4.3,
is shown in Figure~1 \cite{81}.

In the configuration of two parallel plates (like shown in Figure~1) and a sphere above
a plate (up to now, the latter was used in the most precise measurements of the
Casimir interaction) the additional force due to  a symmetron field was calculated in
Ref.~\cite{84}. For a sphere-plate geometry, these calculations were performed under the
conditions $mR\ll 1$, $mR\sim 1$, and $mR\gg 1$, where $m$ is the symmetron mass in the
vacuum  and $R$ is the sphere radius, with account of the screening effects.

\begin{adjustwidth}{-\extralength}{0cm}
\begin{figure}[H]
\vspace*{-25.8cm}
\centerline{\hspace*{4.7cm}
\includegraphics[width=9.5in]{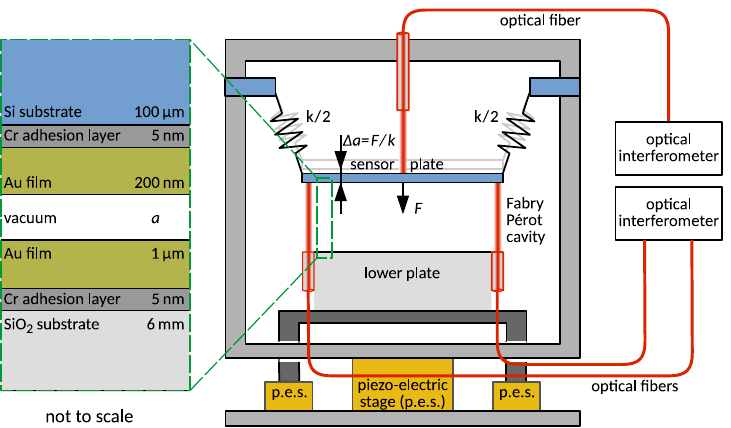}}
\vspace*{0.25cm}
\caption{Schematic of the setup of CANNEX experiment for measuring the Casimir force
between two parallel plates at large separations. The pressure between the fixed lower
plate and the movable upper sensor plate separated by a distance $a$ is measured by
monitoring the extension $\Delta a$ using the optical interferometer. The material
structure of both the lower and upper plates is shown not to scale at the left of the
figure.}
\end{figure}
\end{adjustwidth}

In the case of two parallel plates of area $S$, it was found that at sufficiently small
separations $a$ between them satisfying the condition $am<\pi$ the additional symmetron
force per plate area, i.e., the additional pressure, is given by \cite{84}
\begin{linenomath}
\begin{equation}
P_{\rm s}=\frac{F_{\rm s}}{S}=-\frac{m^4}{4\lambda}.
\label{eq25}
\end{equation}
\end{linenomath}

At larger separations, the symmetron pressure goes to zero exponentially fast. These results
are obtained for sufficiently dense plates with $\varepsilon\gg m^2M^2$, where $M$ is the
mass scale entering the effective potential in Eq.~(\ref{eq17}). This condition allows to
put $\Phi\approx 0$ inside the plates.

When considering the sphere-plate configuration, it was also assumed that these bodies are
sufficiently dense. Under this condition, for the spheres of large radii,
$R\gg m^{-1}$, the following approximate expressions for the additional symmetron force
were obtained \cite{84}
\begin{linenomath}
\begin{eqnarray}
&&
F_{\rm s}=-\frac{m^4}{4\lambda}\pi R^2,\qquad a<\frac{\pi}{m}-R,
\nonumber\\[1mm]
&&
F_{\rm s}=-\frac{m^4}{4\lambda}\pi\left(\frac{\pi}{m}-a\right)\,
\left(2R+a-\frac{\pi}{m}\right),
\qquad \frac{\pi}{m}-R<a<\frac{\pi}{m},
\nonumber\\[1mm]
&&
F_{\rm s}=0,\qquad a>\frac{\pi}{m},
\label{eq26}
\end{eqnarray}
\end{linenomath}
where $a$ is the closest sphere-plate separation.

The approximate analytic calculation of $F_{\rm s}$ is also possible for the spheres of
small radii $R\ll m^{-1}$ above a plate. The result is \cite{84}
\begin{linenomath}
\begin{equation}
F_{\rm s}=-\frac{4\pi m^3R}{\sqrt{2}\lambda}\,{\rm tanh}\frac{m(a+R)}{\sqrt{2}}\,
{\rm sech}^2\frac{m(a+R)}{\sqrt{2}}.
\label{eq27}
\end{equation}
\end{linenomath}

As is seen from Eq.~(\ref{eq27}), in the limiting case $a\to 0$, i.e., when the sphere
approaches the plate, the magnitude of the symmetron force decreases to
\begin{linenomath}
\begin{equation}
|F_{\rm s}|=\frac{2\pi}{\lambda}m^2(mR)^2,
\label{eq28}
\end{equation}
\end{linenomath}
where $mR\ll 1$ in this case.

In the region of intermediate values of the sphere radius $mR\sim 1$, the additional
force due to the symmetron field was computed numerically \cite{84}.

For obtaining constraints on the parameters of a symmetron field, it was suggested
\cite{84} to use a setup similar to that of Ref.~\cite{85}. In the proposed setup,
a sphere of $R=150~\upmu$m radius is spaced at a distance $a=15~\upmu$m from a rotating
disk covered with rectangular trenches of $50~\upmu$m depth in high vacuum.
As a result, the distance between the sphere bottom and the disk surface varies between
$a_{\min}=15~\upmu$m and $a_{\max}=65~\upmu$m. Taking into account that all the known
forces at these separations are much smaller than the experimental error
$\Delta F=0.2~$fN, the constraints on the symmetron force can be obtained from the
inequality \cite{84}
\begin{linenomath}
\begin{equation}
F_{\rm s}(a_{\min})-F_{\rm s}(a_{\max})=\pm \Delta F,
\label{eq29}
\end{equation}
\end{linenomath}
using the expressions for $F_{\rm }$ presented above. The expected constraints which can be
obtained in this way are discussed in Ref.~\cite{84}.

\subsection{Constraints on the Environment-Dependent Dilaton
from Measuring the Casimir Force}

The parameters of the environment-dependent dilaton can be constrained from the same
experiments as the parameters of chameleon and symmetron. Thus, in Refs.~\cite{18,86}
the dilaton parameters were constrained using the experimental data of Ref.~\cite{62}
on measuring the quantum states of ultracold neutrons near a mirror. As discussed in
Section~3, these data have already been used for constraining the parameters of a
chameleon model.

In Ref.~\cite{18} it was also suggested to constrain the parameters of an
environment-dependent dilaton from the CANNEX experiment (see Figure~1) on measuring
the Casimir force between two parallel plates at large separations \cite{81,82,83}.
For this purpose, using the potential (\ref{eq22}) and the coupling function (\ref{eq21}),
the exact solutions for a dilaton field were obtained in the configurations with one and
two mirrors.

The additional dilaton pressure arising between two parallel plates with an effective
area of $1~\mbox{cm}^2$ was computed in Ref.~\cite{86} in application to the CANNEX
experiment. In this experiment, it was assumed that the separation distance between the
plates can be varied from 1.5 to $15~\upmu$m. It was also suggested to vary the
pressure around the plates by admitting Xe gas into the vacuum chamber. This option
allows making the differential force measurements which present many advantages in the
case of an environment-dependent force. As a result, the dilaton field between the
plates and the corresponding additional pressure have been computed numerically under
the condition
\begin{linenomath}
\begin{equation}
4\pi G A_0\Phi^2\ll 1.
\label{eq30}
\end{equation}
\end{linenomath}
This condition ensures that one can omit the coupling to matter of higher orders which
is neglected in Eq.~(\ref{eq21}).

Taking into account the planned sensitivity of the CANNEX experiment to force measurements
equal to $0.1~\mbox{nN/m}^2$, the prospective constraints on the dilaton parameters
$\lambda$ and $A_0$ were obtained in Ref.~\cite{86} from an assumption that no extra
forces in addition to the Casimir one were registered.

\section{Discussion}

In the foregoing, we have considered different models of dark energy,
which makes up to approximately 68\% of the total energy of the Universe.
These models differ significantly in their physical meaning and
theoretical background. In some sense, the model of dark energy using
Einstein's equations with the cosmological term provides the most
economic description of the dark energy which does not require any
changes in the mathematical formalism of fundamental physical theories
and introduction of additional
physical substances with unusual properties. It follows that this
model can be considered as preferable.

All the other types of models considered above, using the concepts of
the quintessence, modified gravity and hypothetical particles with
unusual physical properties, in any event are based on an introduction of
some additional scalar field with one or other type of the interaction
potential and the function describing its interaction with the baryonic
matter. There are many models specifying these functions in the one
way or another, and in each case much work should be done to reconcile
the model properties with all the available data from different
experiments and astrophysical observations.

It should be emphasized that the chameleon, symmetron and
environment-dependent dilaton fields and corresponding particles are
radically different from the particles and fields used in the Standard
Model of elementary particle physics. The particles and fields
introduced for the understanding of dark energy are not similar to
those introduced, for instance, in different approaches to the
theoretical description of dark matter. In fact, the hypothetical
particle constituents of dark matter, such as axions or weakly
interacting massive particles, can be understood as some extensions
of the Standard Model. Axions, for instance, were introduced
\cite{87,88} for a resolution of the problem of strong CP violation
in quantum chromodynamics with no relation to the concept of dark
matter.

It might be well to point out that in the framework of quantum theory
the explanation of dark energy in terms of the cosmological constant
is burdened by the problem called the vacuum catastrophe (see Section
2) and the alternative explanations using a variety of scalar fields
imply a departure from the well approved general theory of relativity
in favor of the scalar-tensor theory. Because of this, one may expect
that the final resolution of the problem of dark energy will be found only
in the context of quantum theory of gravitation. Meantime any
experimental constraints on the proposed models of dark energy are
of much importance by guiding the most prospective ways for further
progress in cosmology.

\section{Conclusions}

To conclude, none of the model approaches to understanding of the dark
energy discussed above can be considered as fully satisfactory. This
increases the role of experiment, which may not only to confirm the
theoretical predictions, but to place so strong constraints on the
parameters of some model that it will become completely unusable. In
this regard, the laboratory experiments are the most promising because
all their parameters are under the strict control which is often not
the case for astrophysical observations.

In the above, we mentioned several laboratory experiments aimed
to constrain the parameters of chameleon, symmetron, and
environment-dependent dilaton fields, such as using the atom
interferometry, interaction of the hypothetical scalar fields with
the electromagnetic field, and scattering of ultracold neutrons (see
Section 4). The main attention, however, was devoted to the possibility
of constraining the parameters of these scalar fields from precise
measurements of the Casimir force.

As is shown in the literature reviewed in Section 4, the parameters of
chameleon, symmetron, and environment-dependent dilaton fields can be
constrained from the experiments on measuring the Casimir force. The
prospective constraints, which can be obtained in this way, are quite
competitive, as compared to the other laboratory experiments. For
obtaining these constraints, it will be necessary, however, to create
the next generation of setups which will allow measuring the Casimir
interaction at large separations up to 10 micrometers and even more.

This work is currently in progress. Its successful completion will
allow not only to place new more strong constraints on the models of
dark energy, but also solve the remaining problems of the Casimir
physics.

\vspace{6pt}

\funding{
This work was partially funded by the
Ministry of Science and Higher Education of Russian Federation
("The World-Class Research Center: Advanced Digital Technologies,"
contract No. 075-15-2022-311 dated April 20, 2022).
The work of
V.M.M.~was also partially carried out in accordance with the Strategic
Academic Leadership Program "Priority 2030" of the Kazan Federal
University.}

\begin{adjustwidth}{-\extralength}{0cm}

\reftitle{References}

\end{adjustwidth}
\end{document}